\documentclass[12pt]{article}

\usepackage{mathrsfs}
\usepackage[T1]{fontenc}
\usepackage{mathpazo}
\usepackage{setspace}
\usepackage{amsfonts}
\usepackage{amssymb}
\usepackage{amsmath}
\usepackage{esint}                   
\usepackage{epsfig}
\usepackage{latexsym}
\usepackage{color}
\usepackage{graphicx}
\usepackage{hyperref}
\usepackage{nicefrac}
\usepackage[latin1]{inputenc}
\usepackage{pstricks}
\usepackage{slashed}
\usepackage{multirow}
\usepackage{ulem}
\usepackage{comment}
\usepackage[nosort]{cite}
\usepackage{datetime}
\usepackage{tikz-cd}
\usepackage{float}

\usepackage{comment}
\usepackage{ytableau}

\usepackage{braket}

\def\hybrid{\topmargin -30pt    \oddsidemargin 0pt 
        \headheight 0pt \headsep 0pt
        \textwidth 6.25in       
        \textheight 9.5in       
        \marginparwidth .875in
        \parskip 5pt plus 1pt   \jot = 1.5ex}

\hybrid

\def\baselinestretch{1.2}

\catcode`\@=11

\def\marginnote#1{}
\newcount\hour
\newcount\minute
\newtoks\amorpm
\hour=\time\divide\hour by60
\minute=\time{\multiply\hour by60 \global\advance\minute by-\hour}
\edef\standardtime{{\ifnum\hour<12 \global\amorpm={am}%
        \else\global\amorpm={pm}\advance\hour by-12 \fi
        \ifnum\hour=0 \hour=12 \fi
        \number\hour:\ifnum\minute<10 0\fi\number\minute\the\amorpm}}
\edef\militarytime{\number\hour:\ifnum\minute<10 0\fi\number\minute}

\def\draftlabel#1{{\@bsphack\if@filesw {\let\thepage\relax
   \xdef\@gtempa{\write\@auxout{\string
      \newlabel{#1}{{\@currentlabel}{\thepage}}}}}\@gtempa
   \if@nobreak \ifvmode\nobreak\fi\fi\fi\@esphack}
        \gdef\@eqnlabel{#1}}
\def\@eqnlabel{}
\def\@vacuum{}
\def\draftmarginnote#1{\marginpar{\raggedright\scriptsize\tt#1}}

\def\draft{\oddsidemargin -.5truein
        \def\@oddfoot{\sl preliminary draft \hfil
        \rm\thepage\hfil\sl\today\quad\militarytime}
        \let\@evenfoot\@oddfoot \overfullrule 3pt
        \let\label=\draftlabel
        \let\marginnote=\draftmarginnote
   \def\@eqnnum{(\theequation)\rlap{\kern\marginparsep\tt\@eqnlabel}%
\global\let\@eqnlabel\@vacuum}  }

\def\draft2{
        \def\@oddfoot{\sl preliminary draft \hfil
        \rm\thepage\hfil\sl\today\quad\militarytime}
        \let\@evenfoot\@oddfoot \overfullrule 3pt
        \let\marginnote=\draftmarginnote
   \def\@eqnnum{(\theequation)\rlap{\kern\marginparsep\tt\@eqnlabel}%
\global\let\@eqnlabel\@vacuum}  }

\def\preprint{\twocolumn\sloppy\flushbottom\parindent 2em
        \leftmargini 2em\leftmarginv .5em\leftmarginvi .5em
        \oddsidemargin -.5in    \evensidemargin -.5in
        \columnsep .4in \footheight 0pt
        \textwidth 10.in        \topmargin  -.4in
        \headheight 12pt \topskip .4in
        \textheight 6.9in \footskip 0pt
        \def\@oddhead{\thepage\hfil\addtocounter{page}{1}\thepage}
        \let\@evenhead\@oddhead \def\@oddfoot{} \def\@evenfoot{} }

\def\numberbysection{\@addtoreset{equation}{section}
        \def\theequation{\thesection.\arabic{equation}}}

\def\underline#1{\relax\ifmmode\@@underline#1\else
        $\@@underline{\hbox{#1}}$\relax\fi}

\def\titlepage{\@restonecolfalse\if@twocolumn\@restonecoltrue\onecolumn
     \else \newpage \fi \thispagestyle{empty}\c@page\z@
        \def\thefootnote{\fnsymbol{footnote}} }

\def\endtitlepage{\if@restonecol\twocolumn \else \newpage \fi
        \def\thefootnote{\arabic{footnote}}
        \setcounter{footnote}{0}}  

\catcode`@=12
\relax

\def\figcap{\section*{Figure Captions\markboth
        {FIGURECAPTIONS}{FIGURECAPTIONS}}\list
        {Figure \arabic{enumi}:\hfill}{\settowidth\labelwidth{Figure
999:}
        \leftmargin\labelwidth
        \advance\leftmargin\labelsep\usecounter{enumi}}}
 \relax
\def\tablecap{\section*{Table Captions\markboth
        {TABLECAPTIONS}{TABLECAPTIONS}}\list
        {Table \arabic{enumi}:\hfill}{\settowidth\labelwidth{Table
999:}
        \leftmargin\labelwidth
        \advance\leftmargin\labelsep\usecounter{enumi}}}
 \relax
\def\reflist{\section*{References\markboth
        {REFLIST}{REFLIST}}\list
        {[\arabic{enumi}]\hfill}{\settowidth\labelwidth{[999]}
        \leftmargin\labelwidth
        \advance\leftmargin\labelsep\usecounter{enumi}}}
 \relax
\makeatletter
\newcounter{pubctr}
\def\publist{\@ifnextchar[{\@publist}{\@@publist}}
\def\@publist[#1]{\list
        {[\arabic{pubctr}]\hfill}{\settowidth\labelwidth{[999]}
        \leftmargin\labelwidth
        \advance\leftmargin\labelsep
        \@nmbrlisttrue\def\@listctr{pubctr}
        \setcounter{pubctr}{#1}\addtocounter{pubctr}{-1}}}
\def\@@publist{\list
        {[\arabic{pubctr}]\hfill}{\settowidth\labelwidth{[999]}
        \leftmargin\labelwidth
        \advance\leftmargin\labelsep
        \@nmbrlisttrue\def\@listctr{pubctr}}}
 \relax
\makeatother

\def\be{\begin{equation}}
\def\ee{\end{equation}}
\def\ba{\begin{eqnarray}}
\def\ea{\end{eqnarray}}

\def\bx {{\bf x}}
\def\bk {{\bf k}}

\def\b{\beta}

\def\d{\delta}

\def\th{\theta}
\def\Th{\Theta}

\def\l{\lambda}

\def\no{\noindent}

\def\qq{\qquad}

\def\IR{\relax{\rm I\kern-.18em R}}

\def\bx {{\bf x}}

\def\inv{^{\raise.0ex\hbox{${\scriptscriptstyle -}$}\kern-.05em 1}}

\def \ha {{\frac{1}{2}}}

\def \ov {\over}

\def\const{{\rm const.}}

\def\half{{\textstyle {1 \over 2}}}

\newcommand{\bb}{\hskip -0.1cm}

\def\tr{\textrm{Tr}}

\begin{document}


\renewcommand{\theequation}{\thesection.\arabic{equation}}
\csname @addtoreset\endcsname{equation}{section}

\begin{titlepage}
\begin{center}

\renewcommand*{\thefootnote}{\arabic{footnote}}

\phantom{xx}
\vskip 0.5in

{\large {\bf Nonabelian ferromagnets with three-body interactions}}

\vskip 0.5in

{\bf Alexios P. Polychronakos$^{1,2}$}\hskip .15cm and \hskip .15cm
{\bf Konstantinos Sfetsos}$^{3}$

\vskip 0.14in

${}^1\!$ Physics Department, City College of New York\\
160 Convent Avenue, New York, NY 10031, USA\\
{\footnotesize{\tt apolychronakos@ccny.cuny.edu}}\\
\vskip 0.3cm
${}^2\!$ The Graduate School and University Center, City University of New York\\
365 Fifth Avenue, New York, NY 10016, USA\\
{\footnotesize{\tt apolychronakos@gc.cuny.edu}}

\vskip .15in

${}^3\!$
Department of Nuclear and Particle Physics, \\
Faculty of Physics, National and Kapodistrian University of Athens, \\
Athens 15784, Greece\\
{\footnotesize{ ksfetsos@phys.uoa.gr}}\\

\vskip .3in
\today

\vskip .2in

\end{center}

\vskip .2in

\centerline{\bf Abstract}

\no
We study the thermodynamics of nonabelian ferromagnets consisting of atoms in the
fundamental representation of $SU(N)$ and 
interacting with two-body and three-body interactions. Using a mean field approach,
we uncover an intricate phase structure, depending
on the relative strength and sign of the two-body and three-body coupling constants.
In the case where two-body interactions
are ferromagnetic and three-body ones are antiferromagnetic, we uncover a rich cascade of
phase transitions, the appearance of phases with two distinct polarization directions being the
most striking novel feature. Our results are relevant to magnetic systems where higher-body
interactions cannot be neglected.

\vskip .4in

\vfill

\end{titlepage}
\vfill
\eject



\def\baselinestretch{1.2}
\baselineskip 20 pt

\newcommand{\eqn}[1]{(\ref{#1})}

\tableofcontents


\section{Introduction}
\label{intro}

\no
Magnetic systems characterized by higher internal $SU(N)$ symmetry have attracted significant attention in both experimental and theoretical research. These systems arise in diverse settings, including ultracold atomic gases \cite{Dud,Ghu,Gor,Zha,Mag,Cap,Mukherjee:2024ffz}, spin chains \cite{Aff,Pola}, lattice-based models of interacting atoms \cite{KT,BSL,RoLa,YSMOF,TK,Totsuka,TK2}, and scenarios involving nonabelian magnetic fields \cite{DY,YM,HM,PSferro}. They display unconventional collective behavior and possess a complex and intriguing phase structure.

In recent work \cite{PSferro,PSlargeN,PSferroN,PShigher} we considered ferromagnets consisting of
atoms on fixed positions with degrees of freedom in an irreducible representation of $SU(N)$
and pairwise interactions.
We derived the thermodynamics of such systems and uncovered an intricate and nontrivial phase
structure with qualitatively new features, including several critical temperatures (vs. only one Curie
temperature for $SU(2)$), metastability, and hysteresis phenomena, both in the temperature and the magnetic field.
In the limit where the rank of the group $N$ scales as the square root
of the number of atoms we uncovered an even more intricate phase structure,
featuring a triple critical point and different temperature scales.

\no
In all previous work, atoms interacted only pairwise. In more realistic situations, atoms can also
interact with higher-body interactions, although of decreasing strength. In this paper we explore the
influence and effects of higher-body interactions by studying the phenomenologically most relevant
case of atoms in the fundamental representation of $SU(N)$ and interacting with pairwise as well as
three-body terms. We were able to perform the thermodynamic analysis explicitly, although its increased
complexity also calls for numerical investigations, and derived equilibrium equations for the states of the 
system. We uncovered a rich pattern of phase structures and transitions depending
on the relative strength and sign of the two-body and three-body coupling constants, with novel
states of broken symmetry appearing compared to the case of pure two-body interactions.
In the most interesting case of competing ferromagnetic two-body and antiferromagnetic three-body
interactions, a cascade of phase transitions occurs at various critical temperatures. For the case of
$SU(3)$, which we analyzed fully, a particular symmetry breaking pattern from high to low temperature is
$SU(3) \to SU(2)\times U(1) \to U(1)\times U(1) \to U(1) \times SU(2)$, where the two
orderings of factors in $SU(2) \times U(1)$ stand for two qualitatively
different states of this symmetry appearing at different temperatures.

\no
The organization of the paper is as follows:
In section \ref{model} we introduce the model, present its free energy in the thermodynamic limit, and derive
its equilibrium equations and the conditions for stability of its configurations.
In section \ref{strong} we derive the properties of the model in the strong ferromagnetic case in which both
two- and three-body interactions are ferromagnetic and make contact with our previous results \cite{PSferro}
when three-body terms vanish. In section \ref{weak} we examine the weakly ferromagnetic case in which
two-body terms are antiferromagnetic and note its deviations from the standard nonabelian ferromagnet.
In section \ref{mixed} we analyze the most interesting and nontrivial case where two-body interactions are
ferromagnetic and three-body ones are antiferromagnetic, and, focusing on the case of $SU(3)$, derive its intricate
phase transition structure in various domains of the couplings. Finally, in section \ref{conclusions} we
present our conclusions and point to possible future directions of investigation.

\section{The model and the thermodynamic limit}
\label{model}

We consider a set of $n$ "atoms" on fixed lattice positions, each having $N$ internal states and interacting with
few-body interactions. The set of fundamental $SU(N)$ operators $j^a_i$, $a=1,\dots, N^2-1$, acting on the states of
atom $i$, plus the identity operator constitute a complete basis of operators on the Hilbert space of the system.
Assuming that all interactions are invariant under a common change of basis of the $N$ states of each atom, the
$m$-body interaction between atoms $i_1\dots,i_m$ can be written, up to an irrelevant additive constant, as
\be
H_{m;i_1,\dots,i_n} = c_{i_1,\dots,i_m}\sum_{a_i ,\dots, a_m = 1}^{N^2 -1} t_{a_1,\dots, a_m} \, j_{i_1}^{\,a_1} \cdots j_{i_m}^{\,a_m} \ ,
\label{Hnij}
\ee
where $t_{a_1,\dots,a_m}$ is an $SU(N)$-invariant $m$-tensor and $c_{i_1,\dots,i_m}$ coupling constants.
For two-body interactions, $t_{a_1,a_2} = \delta_{a_1,a_2}$. For three-body interactions there are two invariant
tensors: the antisymmetric one involving the coupling constant $f_{a_1,a_2,a_3}$, and the symmetric one
$d_{a_1,a_2,a_3}$ (sometimes called the "anomaly" in particle physics contexts), expressible in terms of the
$N$-dimensional fundamental generators $F^a$ of $SU(N)$ as
\be
d_{a_1,a_2,a_3} + i f_{a_1,a_2,a_3} = 4\tr ( F^{a_1} F^{a_2} F^{a_3} )\ , \quad \text{assuming} \quad
\tr ( F^a F^b ) = \half \delta_{a b}\ .
\ee
For higher interactions there are several tensors. For instace, for four-body interactions there are two reducible tensors,
$\delta_{a_1,a_2} \delta_{a_3,a_4}$
and $\delta_{a_1,a_3} \delta_{a_2,a_4}$, and six irreducible ones, $\tr (F^{a_1} F^{a_2} F^{a_3} F^{a_4})$ and its
non-cyclic permutations of $a_1,\dots, a_4$.

\no
We will focus on the lowest nontrivial case of three-body interactions, expecting higher-order ones to be of diminishing
strength and importance. The full Hamiltonian becomes
\be
H = \sum_{i,j=1}^n \sum_{a=1}^{N^2-1} c_{ij} \ j_i^a j_j^a + \sum_{i,j,k=1}^n \sum_{a,b,c=1}^{N^2-1}
\big( c_{ijk}\ d_{abc} + {\tilde c}_{ijk}\ f_{abc} \big)\ j_i^a j_j^b j_k^c\ ,
\ee
where the couplings $c_{ij}$ and $c_{ijk}$ are fully symmetric in their indices while ${\tilde c}_{ijk}$ is fully antisymmetric.
We also take the couplings to vanish when any two of their indices coincide to eliminate self-interactions, which lead to
trivial constants in the two-body part and to two-body terms in the three-body part that can be absorbed in the
two-body part.

\no
Reasonable physical assumptions restrict the form of the couplings. We assume that the system is homogeneous,
that is, the couplings are translationally invariant under a shift of all the atoms by the same lattice translation (as long
as we are not close to the boundary of the lattice). In terms of the lattice positions of the atoms $\vec r$,
\be
c_{{\vec r} , {\vec s}} = c_{{\vec r} - {\vec s}} \ ,\quad c_{{\vec r},{\vec s},{\vec q}} = 
c_{{\vec r}-{\vec s},{\vec s}-{\vec q}}\ ,\quad {\tilde c}_{{\vec r},{\vec s},{\vec q}} = 
{\tilde c}_{{\vec r}-{\vec s},{\vec s}-{\vec q}}  \ .
\ee
The symmetries of $c_{ij},c_{ijk}$ and ${\tilde c}_{ijk}$ imply
\be
c_{-{\vec r}} = c_{\vec r} \ ,\quad c_{{\vec r},{\vec s}} = c_{{\vec s},{\vec r}} = c_{-{\vec r},{\vec s}} \ ,
\quad {\tilde c}_{{\vec r},{\vec s}} = -{\tilde c}_{{\vec s},{\vec r}} = -{\tilde c}_{-{\vec r},{\vec s}} \ .
\ee
Therefore, each atom couples to fixed weighted averages of the $SU(N)$ generators of its neighboring
atoms. The Hamiltonian becomes
\be
\begin{split}
H = \sum_{a=1}^{N^2-1} \sum_{\vec r} j_{\vec r}^a \sum_{\vec s}c_{\vec s}\ j_{{\vec r}+{\vec s}}^a 
&+  \sum_{a,b,c=1}^{N^2-1} d_{abc} \sum_{\vec r} j_{\vec r}^a \sum_{{\vec s}} j_{{\vec r}+{\vec s}}^b \sum_{{\vec q}} 
c_{{\vec s},{\vec q}} \ j_{{\vec r}+{\vec s}+{\vec q}}^c\cr
&+  \sum_{a,b,c=1}^{N^2-1} f_{abc} \sum_{\vec r} j_{\vec r}^a \sum_{{\vec s}} j_{{\vec r}+{\vec s}}^b \sum_{{\vec q}} 
{\tilde c}_{{\vec s},{\vec q}} \ j_{{\vec r}+{\vec s}+{\vec q}}^c\  .
\end{split}
\ee
\no
We now make the assumption that all interactions are reasonably long-range, that is, each atom couples to
several of its neigboring atoms. This justifies the mean field condition that, in the thermodynamic limit,
the weighted average of the neighboring atoms is well approximated by their average over the full
lattice.\footnote{The validity of the mean field approximation is
strongest in three dimensions, since every atom has a higher number of near neighbors and the statistical
fluctuations of their averaged coupling are weaker, but is expected to hold also in lower dimensions.}
That is,
\be
\sum_{\vec s} c_{\vec s}\, j_{{\vec r}+{\vec s}}^a \simeq \Bigl(\sum_{{\vec s}} c_{\vec s} \Bigr)\,
{1\over n} \sum_{i} j_{i}^a = -{c \over n} J^a\ ,
\label{meanfield}\ee
where we defined the total $SU(N)$ generators $J^a$ and the effective mean coupling $c$
\be
J^a = \sum_{i} j_i^a \ , \qq c = - \sum_{{\vec s}} c_{\vec s}\ .
\ee
The minus sign is introduced such that ferromagnetic interactions, driving atom states to align,
correspond to positive $c$. 
Similarly
\be
\sum_{\vec q} c_{{\vec s},{\vec q}}\, j_{{\vec r}+{\vec s}+{\vec q}}^a \simeq 
\Bigl(\sum_{{\vec q}} c_{{\vec s},{\vec q}} \Bigr)\,
{1\over n} \sum_{i} j_{i}^a = {1\over n} \Bigl(\sum_{{\vec q}} c_{{\vec s},{\vec q}} \Bigr) J^a
\label{meanfield31}\ee
and
\be
\sum_{\vec s} \Bigl(\sum_{{\vec q}} c_{{\vec s},{\vec q}} \Bigr) \, j_{{\vec r}+{\vec s}}^a \simeq 
\Bigl(\sum_{{\vec s},{\vec q}} c_{{\vec s},{\vec q}} \Bigr)\,
{1\over n} \sum_{i} j_{i}^a = -{q \over n} J^a\ ,
\label{meanfield32}\ee
with
\be
q = -\sum_{{\vec s},{\vec q}} c_{{\vec s},{\vec q}} \ .
\label{meanfield33}\ee
Finally, the corresponding equations (\ref{meanfield31},\ref{meanfield32},\ref{meanfield33}) for ${\tilde c}_{{\vec s},{\vec q}}$
lead to ${\tilde q} = 0$, due to the antisymmetricity properties of ${\tilde c}_{{\vec s},{\vec q}}$. Under the above mean field
conditions, the two-body part of the Hamiltonian assumes the form
\be
\sum_{a=1}^{N^2-1} \sum_{{\vec r}}  j_{\vec r}^a \sum_{{\vec s}} c_{\vec s} \  j_{{\vec r}+{\vec s}}^a  \, 
  \simeq  -{c \over n} \sum_{a=1}^{N^2 -1} \sum_{\vec r} j_{\vec r}^a  J_a
= -{c \over n} \sum_{a=1}^{N^2 -1} J_a^2  = -{c\ov n} C^{(2)}\ .
\label{meanfield2}\ee
So the two-body part of $H$ is proportional to the quadratic Casimir of the total $SU(N)$ group
$C^{(2)}$. Similarly, the three-body part becomes approximately
\be
-{q \over n^2} \sum_{abc} d_{abc} J^a J^b J^c = -{q\over n^2} C^{(3)}
\label{meanfield3}\ee
and is proportional to the cubic Casimir $C^{(3)}$ of the total $SU(N)$ group. The
antisymmetric part $f_{abc}$ does not contribute in the mean field approximation. Note also the ${1\ov n}$
and $1\ov n^2$ scaling factors in the two- and three-body parts respectively, with $n$ the number of atoms,
which will be crucial for the proper thermodynamic limit of the system.

\no
We conclude that the full effective interaction contains the quadratic and cubic Casimirs of the total $SU(N)$ generators
and involves two effective coupling constants,
\be
H = -{c \over n} C^{(2)} - {q\ov n^2} C^{(3)}\ .
\label{H0}\ee
To proceed, we need an efficient way to calculate these Casimirs for each irreducible component of the
total $SU(N)$ group in the Hilbert space of the system and to evaluate the multiplicity of each irrep in the Hilbert space.
This can most efficiently be done in the momentum representation of \cite{PScompo}, which we briefly review here.

\no
Irreducible representations (irreps) of $SU(N)$ arising in the decomposition of $n$ fundamentals can be
parametrized in terms of a set of nonnegative ordered distinct integers $k_1 > k_2 > \cdots > k_N$,
collectively denoted $\bk$, satisfying the constraint
\be
{\sum_{i=1}^N k_i }= n +{N(N-1)\over 2}\ .
\label{lengyt}
\ee
The $k_i$ are related to the lengths of rows $\ell_i$ of the corresponding Young tableau as $
\ell_i = k_i - k_N + i-N$.
The Casimirs in \eqn{H0}, up to terms involving lower Casimirs that can be absorbed in redefinitions
of the coefficient $c$, can be expressed in terms of $\bk$ as
\be
C^{(2)} = \ha \sum_{i=1}^N k_i^2 + \const\ ,\qq C^{(3)} = {1\ov 6} \sum_{i=1}^N k_i^3 + \const\ ,
\ee
The partition function of the system then becomes
\be
Z = \tr\ e^{-\b H} = \sum_{\bk}\, \dim ({\bf k})\, d_{n;{\bf k}} \, e^{-\b H} = \sum_{\bk} e^{-\b \, {\bf F}}\ .
\ee
In the above, $d_{n;{\bf k}}$ accounts for the number of irreps $d_{n;{\bf k}}$ in the tensor product
$(\otimes F)^n$ of the fundamental irreps carried by the atoms, and $\dim ({\bf k})$
for the number of states in each irrep, and are given by \cite{PScompo}
\be
\label{dimd}
\dim ({\bf k}) =  \prod_{j>i=1}^N {k_i - k_j  \over j-i} \ ,  \qq
 d_{n;{\bf k}} = n!\, {\displaystyle \prod_{j>i=1}^N (k_i - k_j) \over \displaystyle \prod_{i=1}^N k_i !}\ .
\ee
In the thermodynamic limit, $n \gg 1$ and $k_i \sim n \gg 1$\footnote{From the constraint \eqn{lengyt} we see that at least some of the $k_i$ must scale like $n$. Configurations where some of the $k_i$ do not scale like $n$ contribute
subleadingly in \eqn{dimd}.}. Using the Stirling formula and ignoring
terms subleading in $n$ and trivial constants, the free energy 
$\bf F$ becomes
\be
\label{Ffull}
{\bf F}(T,\bk)= \sum_{i=1}^N \bigg(T k_i \ln k_i - {c\ov 2n} k_i^2 - {q\ov 6n^2} k_i^3\bigg)\ .
\ee
Note that among the factors in \eqn{dimd} only $ \displaystyle \prod_{i=1}^N k_i !$ contributes in the
thermodynamic limit.
We now introduce the intensive variables $x_i = k_i /n$ and, for later
convenience, the rescaled parameters $T_0$ and $\Theta_0$ as
\be
k_i = n x_i \ ,\qq c= N T_0\ , \qq q= N^2\Theta_0\ .
\ee
The $x_i$ represent the average $SU(N)$ polarization per atom and
satisfy the constraint
\be
\label{lengyt1}
\sum_{i=1}^N x_i = 1\  ,
\ee
while the free energy per atom $F = {\bf F}/n$ becomes, up to an irrelevant constant,
\be
\label{Fgen}
F(T,\bx)= \sum_{i=1}^N \bigg(T x_i \ln x_i - {NT_0\ov 2} x_i^2 - {N^2\Th_0\ov 6} x_i^3\bigg)\ .  
\ee
The saddle points of $F(T,\bx)$ determine the thermodynamic equilibrium conditions.
Implementing the constraint \eqn{lengyt1} with a Lagrange multiplier $\l$, these become
\be
\label{jfghj}
f(x_i) = \l\ , \quad \text{where}\quad f(x) = T \ln x- NT_0 x -{N^2\ov 2} \Theta_0 x^2\ .
\ee
All $x_i$ obey the same equation, and are coupled only through the constraint and
the Lagrange multiplier.

\begin{figure} [th!] 
\begin{center}
\includegraphics[height= 6.5 cm, angle=0]{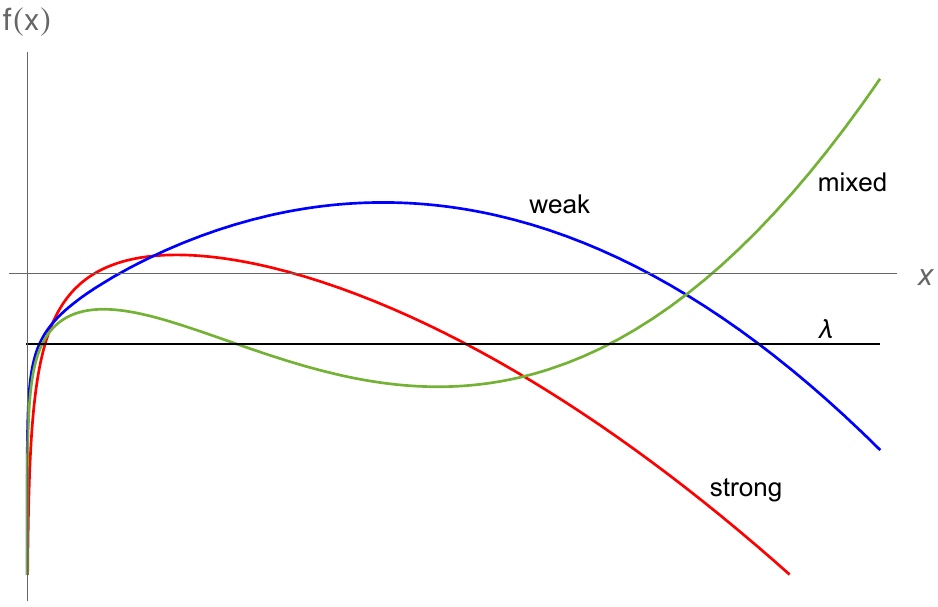}
\end{center}
\vskip -.6 cm
\caption{\small{The function $f(x)$ for the strong (red), weak (}blue) and mixed (green) ferromagnetic
cases. The temperature and value of $\lambda$ are such that $f(x) =\lambda$ have two solutions
for the first two and three solutions for the last one.}
\label{3kaboures}
\end{figure}

\no
The state of the system is determined by the number of $x_i$
at each of the solutions of the equation $f(x)=\l$.
The singlet representation, corresponding to $x_i={1\ov N}$, is always a solution
of \eqn{lengyt1} and \eqn{jfghj} with $\l = f({1\ov N})$. The shape of the function $f(x)$ determines the
existence of additional solutions,
as well as the stability of the corresponding configurations (see fig.~\ref{3kaboures}).
 Qualitatively distinct cases arise depending
on the signs of $T_0$ and $\Theta_0$. In increasing level of complexity, these are:

\vskip -.3 cm

\no
\underline{The paramagnetic case $T_0<0$ and $\Theta_0<0$:} $f(x)$
is an increasing function of $x$ and the only solution is the singlet representation. This is
a trivial case.

\no
\underline{The strongly ferromagnetic case $T_0>0$ and $\Theta_0>0$:} $f(x)$ is concave with
a maximum at $x=x_0$,
\be
\label{xo}
x_0 = {T_0\ov 2 N \Theta_0}\bigg(-1+\sqrt{1+{4 T\Theta_0\ov T_0^2}}\bigg)\ ,
\ee 
and $f(x)=\l$ generically has two solutions.

\no
\underline{The weakly ferromagnetic case $T_0<0$ and $\Theta_0>0$:}  $f(x)$ is concave with
a maximum at $x=x_0$,
\be
x_0 = -{T_0\ov 2 N \Theta_0}\bigg(1+\sqrt{1+{4 T\Theta_0\ov T_0^2}}\bigg)\ ,
\ee 
and, similarly to the previous case, $f(x)=\l$ generically has two solutions.

\no
\underline{The mixed case $T_0>0$ and $\Theta_0<0$:} if $T> -{T_0^2\ov 4 \Theta_0}$,
$f(x)$ is an increasing function of $x$ and the only solution is the singlet. If, however,
$T< -{T_0^2\ov 4 \Theta_0}$, $f(x)$ has a maximum at $x=x_-$ and a minimum at $x=x_+$,
\be
\label{xpxm}
x_\pm  =- {T_0\ov 2 N \Theta_0}\bigg(1\pm \sqrt{1+{4 T\Theta_0\ov T_0^2}}\bigg)\ ,\quad x_-<x_+ \ .
\ee
Hence, depending on the value of $\l$, $f(x)=\l$ has either one or three solutions.
This is the most interesting and nontrivial case.

\no
For the two ferromagnetic cases in which $f(x)=\l$ has two solutions, the $x_i$ will group into two sets,
$p$ of them equal to one of the solutions and $N-p$ equal to the other. Taking into account the
constraint \eqn{lengyt1}, they can be expressed in terms of a single parameter $x$ as (the discussion here parallels that in 
\cite{PSferro} in which $\Th_0=0$)
\be
\label{xixi1}
\begin{split}
& x_i = {1+x\ov N}\ ,\qq i=1,2,\dots, p\ ,
\\
&
x_i = {1-a x\ov N}\ , \qq i=p+1,2,\dots, N\ ,\qq a={p\ov N-p}\ .
\end{split} 
\ee
with $x=0$ corresponding to the singlet. The free energy \eqn{Fgen} in terms of $x$ becomes
\be
\label{F2p}
\begin{split}
F(T,x) = &  {T\ov 1+a} \Big(a(1+x)\ln(1+x)+(1-a x)\ln (1-a x)\Big) 
\\
& - {a\ov 2} (T_0+\Theta_0)x^2 - {a(1-a)\ov 6} \Theta_0 x^3 -T\ln N-{T_0\ov 2} - {\Th_0\ov 6} 
\end{split}
\ee
and equations \eqn{jfghj} reduce to the single equation
\be
\label{gh}
T \ln {1+x\ov 1-a x} - (1+a) x \Big(T_0 +\Theta_0 \big(1+{1-a\ov 2} x\big)\Big) = 0 \ .
\ee
The above equation is obviously solved by the singlet, $x=0$.
Both \eqn{F2p} and \eqn{gh} are invariant under the symmetry
\be
\label{symaa}
a\to 1/a \ ,\qquad  x\to -a x \ ,
\ee
which corresponds to inverting the order of the $x_i$. We may thus restrict our attention to solutions with 
\be
\label{akmeros}
p=0,1,2,\dots ,\Big[{N\ov 2}\Big]\ , \quad -1<x<{1\ov a}\ ,
\ee
where $[\, \cdot \,]$ stands for the integer part.

\no
 Thermodynamically stable configurations correspond to local minima of $F(T,x_i)$.
 Their stability is determined by the positivity of the matrix 
\be
\label{fluuc}
\d^2 F = \ha \sum_{i=1}^N C_i^{-1} \d x_i^2\ ,
\ee
where
\be
C_i^{-1} = f' (x_i ) = {T\ov x_i}- NT_0 -N^2 \Theta_0 x_i \ , \qq i=1,2,\dots , N\ ,
\ee
while also taking into account the constraint $\sum_i \d x_i =0$.
As shown in \cite{PSferro}, stability is ensured if
\be
\label{cci}
\begin{split}
&  C_i>0\ ,\quad i=1,2,\dots , N\ , \quad \text{or}
\\
& C_1< 0\ ,\quad C_i >0 \, ,\quad   i=2,\dots , N\ ,\qq \sum_{i=1}^N C_i <0\ .
\end{split}
\ee
In the case with two solutions and $x_i$ given by \eqn{xixi1}, the coefficients $C_i$ take the form
\be
\label{cici}
\begin{split}
& C_i^{-1} = N \bigg( {T\ov 1+x} - T_0 -\Th_0 (1+x)\bigg) \, ,\quad i=1,2,\dots, p\ ,
\\
& C_i^{-1} =N \bigg( {T\ov 1-a x} - T_0 -\Th_0 (1-ax)\bigg) \,  ,\quad i=p+1,\dots, N\ .
\end{split}
\ee

\no
In what follows we analyze the nontrivial ferromagnetic and mixed cases.
It will also be convenient to define new dimensionless positive parameters $\th$ and $t$ as
\be
\label{thta}
 \th= \bigg|{\Th_0\ov T_0}\bigg |  \ ,\qq t = {|\Th_0|\ov T_0^2} T = \th {T\ov |T_0|} \ .
\ee
The parameter $\th$ measures the relative strength of the cubic interaction compared
to the quadratic one and constitutes the only relevant parameter of the system. As we
shall see, it plays a crucial r\^ole in the structure of the phase structure of the model,
especially when the two interactions have opposite sign.

\section{The strongly ferromagnetic case  $T_0>0$, $\Theta_0>0$}
\label{strong}

In this section we will show that this case
qualitatively resembles that with $\Th_0=0$ in which only the quadratic Casimir is present \cite{PSferro}.

\no
In this case, $f(x)=\l$ generically has two solutions. Using the parametrization \eqn{xixi1}
and the dimensionless parameters \eqn{thta}, \eqn{gh} becomes
\be
\label{gh2a}
t \ln  {1+x\ov 1-a x} -\th (1+a) x \Big(1 +\th \big(1+{1-a\ov 2} x\big)\Big) = 0 \ .
\ee
Apart from the singlet solution $x=0$, \eqn{gh2a} can have solutions with $x>0$ and $x<0$.

\no
The stability coefficients \eqn{cici} become
\be
\label{cicia}
\begin{split}
& C_i^{-1} = N T_0 \bigg( {t \ov \th( 1+ x)}-1 - \th (1+x)\bigg) 
 \, ,\quad i=1,2,\dots, p\ ,
\\
& C_i^{-1} = N T_0 \bigg( {t \ov \th( 1-a  x)}-1 - \th (1-a x)\bigg)  
 \,  ,\quad i=p+1,2,\dots, N\ .
\end{split}
\ee
For the singlet, $C_i^{-1} = N T_0 (t/\th-1-\th)\, \  \text{for all\ }i$. Hence, 
\be
\label{stabsina}
 \text{Singlet: \quad stable for} \quad t> t_s = \th(1+\th)\, ,~~
\text{unstable for} \quad t< t_s\  .
\ee
For non-singlet solutions with two distinct sets of $x_i$, concavity of $f(x_i)$ implies that
$f' (x_i\bb<\bb x_0) >0$ and $f' (x_i\bb>\bb x_0) <0$. Therefore, the stability conditions \eqn{cci} imply that only one of the
$x_i$ can be at the larger solution, where $C_i <0$. Therefore, only solutions with $p=1$ (that is,
$a={1\ov N-1}$) and $x>0$
should be considered, corresponding to fully symmetric one-row configurations.
In this case, the additional condition $\sum_{i=1}^N C_i <0$ in \eqn{cci} gives
\be
\label{aaa}
\begin{split}
&p\Big[{t \ov \th( 1+ x)}-1 - \th (1+x)\Big]^{-1} + (N-p) \Big[{t \ov \th( 1-ax)}-1 - \th (1-ax) \Big]^{-1}<0 \\
&\qq\Longrightarrow\quad t > \th (1+x)(1-a x)\big(1+\th +\th(1-a)x\big)\ .
\end{split}
\ee
There is a critical temperature $t_c$ at which the two positive solutions of \eqn{gh2a} merge and disappear. This is found by setting the derivative 
of the left hand side of \eqn{gh2a} to vanish, leading to the additional equation
\be
\label{tcgh}
t = \th (1+x)(1-a x)\big(1+\th +\th(1-a)x\big)\ , \quad a={1\ov N-1} \ .
\ee
 The values $x=x_c$ and $t=t_c$ satisfying the system of equations (\ref{gh2a} and \ref{tcgh}) determine the critical
 temperature $t_c$ and the critical magnetization $x_c$. Elimination of one of $x_c , t_c$ leads to a
 transcendental equation for the other that can only be solved numerically. Nevertheless, it is easy to show that
 $t_c > t_s$.
 For $t>t_c$ the only solution of \eqn{gh2a} is the singlet, $x=0$. For $t<t_c$ there are two additional
 solutions, an unstable one for $x < x_c$ and a stable one for $x>x_c$.
 The condition \eqn{tcgh} at the critical point
 saturates the stability inequality in \eqn{aaa}, making the critical configuration
 $x_c$ marginally stable, which is a generic phenomenon. For $t<t_s$ the singlet becomes unstable,
 leaving the solution with $x>0$ as the only stable configuration.

\no
At temperatures $t_s <t <t_c$ the singlet and one-row states coexist and are stable, one of them
absolutely stable and the other one metastable. At some intermediate temperature $t_m$ their
free energies will be equal, marking a metastability transition: for $t_m<t<t_c$ the singlet is stable and the
one-row is metastable, whereas for $t_s<t<t_m$ their roles are reversed. The free energy  is 
\be
\label{ftp2}
\begin{split}
{F(t)\ov T_0} = &  {t\ov \th (1+a)} \big(a(1+x)\ln(1+x)+(1-a x)\ln (1-a x)\big) 
\\
& - {a\ov 2} (1+\theta)x^2 - {a(1-a)\ov 6} \theta x^3 -{t\ov \th} \ln N-{1\ov 2} - {\th\ov 6} \ ,
\end{split}
\ee
with $a={1\ov N-1}$ and $x$ determined by  \eqn{gh2a} as a function of $t$.
For the singlet 
\be
{F_{\rm singlet}(t)\ov T_0} = -{t\ov \th} \ln N-{1\ov 2} - {\th \ov 6} \ ,
\label{singlF}\ee
and the metastability transition temperature $t_m$ is determined by $F(t_m) = F_{\rm singlet}(t)$.
Unlike the case $\th = 0$, where $t_m$ can be found analytically \cite{PSferro},
$t_m$ for $\th \neq 0$ is the solution of a transcendental equation. The metastability transition is
of first order, since the free energy is continuous but its temperature derivative is not.
Indeed, from (\ref{ftp2} and \ref{singlF}),
\be
{\partial F (x)\ov\partial t} - {\partial F_{\rm singlet}\ov \partial t} = 
{1\ov \th(1+a)} \Big( a(1+x)\ln(1+x) + (1-ax) \ln(1-ax) \Big) \ ,
\ee
which for stable, $x>0$ configurations, is positive and vanishes only
at $x = 0$.

\no
At low temperatures the solution of \eqn{gh2a} is
\be
\label{xtsmall}
x=(N-1) \Big(1 - N e^{-b/t}+ \dots \Big)\ ,\qq b = \th N \Big(1 +{\th\ov 2} N \Big)\ .
\ee
As $t$ approaches zero, $x$ approaches its maximal value $x=N-1$ corresponding to the
maximally polarized one-row state $x_1 = 1$, $x_2 = \dots = x_N = 0$. 
For $t\to 0$ we have
\be
t\simeq 0:\qq {F\ov T_0} \simeq - {N\ov 2} - {N^2\ov 6} \theta \ .
\ee
The phases of the strongly ferromagnetic case are summarized in table \ref{table1}.
\begin{table}[!ht]
\begin{center}
\begin{tabular}{|c|c|c|c|c|} \hline
  irrep & $t<t_s$ & $t_s<t<t_m $ & $t_m<t<t_c$ & $ t_c<t$
  \\ \hline \hline
{\rm singlet}  &  {\rm unstable}      &  {\rm metastable}   &  {\rm stable }  & {\rm stable}     
\\ \hline
 {\rm 1-row} & {\rm stable}    &  {\rm stable} &   {\rm metastable}    &  {$\times$}     
 \\ \hline
\end{tabular}
\end{center}
\vskip -.4 cm
\caption{\small{Phases in various temperature ranges for $N\geqslant 3$ and their stability.}}
\label{table1}
\end{table}
\no
Qualitatively, 
this case resembles the one with  $\Th_0=0$ in which three-body interactions are
absent \cite{PSferro}. In fact, the corresponding tables for the phase structures are
essentially identical. Hence, we have the following spontaneous symmetry
breaking pattern, from higher to lower temperatures
\be
\label{symmbrea1}
SU(N) \to SU(N-1) \times U(1)\ .
\ee
We may present the phase flow in a simplified pictorial manner as in table \ref{table1a}, in which states are arranged
vertically according to the value of their free energy and unstable states are omitted; solid dot stands for the singlet.
\begin{table}[!ht]
\begin{center}
\begin{tabular}{|c|c|c|c|c|} \hline
$t<t_s$ & $t_s<t<t_m$ & $t_m<t<t_c$ & $ t_c<t$
  \\ \hline \hline
     &  {\rm $\bullet$}   &  \begin{tikzpicture}[scale=.3]\draw[thin] (0,0.4)--(0.01,0.4);
\draw[thick] (0,-0.1)--(3,-0.1);
\draw[thick] (0,-1.1)--(3,-1.1);
\draw[thick] (0,-0.1)--(0,-1.1);\draw[thick] (1,-0.1)--(1,-1.1);\draw[thick] (2,-0.1)--(2,-1.1);\draw[thick] (3,-0.1)--(3,-1.1);
\end{tikzpicture} &    
\\ 
 \begin{tikzpicture}[scale=.3]\draw[thin] (0,-1.4)--(0.01,-1.4);
\draw[thick] (0,-0.1)--(5,-0.1);
\draw[thick] (0,-1.1)--(5,-1.1);
\draw[thick] (0,-0.1)--(0,-1.1);\draw[thick] (1,-0.1)--(1,-1.1);\draw[thick] (2,-0.1)--(2,-1.1);\draw[thick] (3,-0.1)--(3,-1.1);
\draw[thick] (4,-0.1)--(4,-1.1);\draw[thick] (5,-0.1)--(5,-1.1);
\end{tikzpicture}  & \begin{tikzpicture}[scale=.3]
\draw[thick] (0,-0.1)--(4,-0.1);
\draw[thick] (0,-1.1)--(4,-1.1);
\draw[thick] (0,-0.1)--(0,-1.1);\draw[thick] (1,-0.1)--(1,-1.1);\draw[thick] (2,-0.1)--(2,-1.1);\draw[thick] (3,-0.1)--(3,-1.1);\draw[thick] (4,-0.1)--(4,-1.1);
\end{tikzpicture}   &   {\rm $\bullet$}    
& {\rm $\bullet$}    
 \\ \hline
\end{tabular}
\end{center}
\vskip -.4 cm
\caption{\small{Phases in various temperature ranges for $N\geqslant 3$ stacked according to their free energies from
lower to higher values. 
Unstable states, or states that disappear as solutions in a given temperature range, are not marked in the table, and
solid dot stands for the singlet.
The length of the YT for the one-row state increases with lowering temperature.}}
\label{table1a}
\end{table}

\no
As we shall see in the following sections, making one of the parameters $T_0$ or $\Th_0$
negative realizes more interesting cases.

\section{The weakly ferromagnetic case  $T_0<0$, $\Theta_0>0$}
\label{weak}

The analysis runs much along the lines of that for the strongly ferromagnetic case, most
formulae simply carrying over upon changing $\th \to - \th$. There are, nevertheless,
some important qualitative differences in the results depending on the parameter $\th$, making the
system's phase structure approach either a ferromagnetic or a paramagnetic one. Indeed, since $T_0<0$,
we expect that as $\th$ decreases the behavior of the system will become progressively more paramagnetic.
For this reason, we call this phase weakly ferromagnetic.

\no
Since $f(x)$ is concave, $f(x_i)=\l$ has again two solutions. As in the previous case, stability considerations
require that at most one of the $x_i$ be at the larger solution.
Using the parametrization \eqn{xixi1} with $p=1$, we obtain the equilibrium equation
\be
\label{gh2aq}
t \ln {1+x\ov 1-a x} + \th(1+a) x \Big(1 - \th \big(1+{1-a\ov 2} x\big)\Big) = 0 \ ,
\ee
with $a={1\ov N-1}$ and $x>0$. Unlike, however, the previous case, this equation can have
nontrivial stable solutions $x> 0$ at sufficiently low $t$ only for large enough $\th$:
\be
\text{solutions with~}x>0:\qq \th >  {2 \ov N}\ .
\ee
The stability coefficients in \eqn{cici} take the form 
\be
\label{cicib}
\begin{split}
& C_1^{-1} = N |T_0| \Big( {t \ov \th( 1+ x)} +1 - \th (1+x)\Big)  \ ,
\\
& C_i^{-1} =N |T_0| \Big( {t \ov \th( 1-ax)} +1 - \th (1-ax)\Big) \ , \quad a={1\ov N-1} ,\quad i=2,\dots, N\ .
\end{split}
\ee
For the singlet, $x=0$ and $C_i^{-1} =N |T_0| (t/\th + 1- \th)$ for all $i$. Hence, 
\be
\label{stabsinqe}
\begin{split}
\text{Singlet:}\quad& \th<1:\quad \text{stable for all $t$}.
\\
& \th >  1:\quad {\rm stable\ for}\  t>t_s = \th(\th-1)\,,\ \  {\rm unstable\ for}\ t< t_s \ .
\end{split}
\ee
The singlet's stability down to zero temperature is a qualitatively new feature.

\no
For $\th >{2\ov N}$, where a solution of \eqn{gh2aq} with $x>0$ can exist, there is a critical temperature
$t_c$ above which this solution disappears, as in the previous case.
This is given by solving the system of \eqn{gh2aq} and 
\be
\label{tcghw}
t = \th (1+x)(1-a x)\Big(-1+\th \big(1+(1-a)x\big)\Big)\ , \quad a={1\ov N-1} \ .
\ee
The solution still satisfies
$t_c > t_s$ (with $t_s$ being zero for $\th <1$).
For $t>t_c$ the singlet is the only stable state, while for $t<t_c$ there is an additional stable one-row state.
As before, at the critical point the stability condition is saturated, making the critical configuration
 $x_c$ marginally stable. If $t_s>0$, for $t<t_s$ the singlet becomes unstable,
 leaving the one-row solution as the only stable state, while for $t_s =0$ the singlet is always stable.

\no
At low temperatures the solution of \eqn{gh2aq} is
\be
\begin{split}
\label{xtsmalq}
& \th > {2 \ov N}:\quad  x=(N-1) \Big(1 - N e^{-b/t}+ \dots \Big)\ ,\quad b =\th N \Big(\ha \th N -1 \Big)\ ,
 \\
& \th < {2 \ov N}:\quad  x = 0 \ .
\end{split}
\ee
For $\th > {2 \ov N}$ and $t\to 0$,
\be
t\simeq 0:\qq {F\ov |T_0|} \simeq   {N\ov 2} - {N^2\ov 6} \theta \ .
\ee

\no
Metastability between stable states is resolved by examining the free energy
\be
\label{ftp22}
\begin{split}
{F(t,x)\ov |T_0|} = &  {t\ov \th (1+a)} \big(a(1+x)\ln(1+x)+(1-a x)\ln (1-a x)\big) 
\\
& - {a\ov 2} (\theta-1)x^2 - {a(1-a)\ov 6} \theta x^3 -{t\ov \th} \ln N + {1\ov 2} - {\th\ov 6} \ ,
\end{split}
\ee
again with $a={1\ov N-1}$ and $x$ determined by  \eqn{gh2aq} as a function of $t$.
For the singlet,
\be
{F_{\rm singlet}(t)\ov |T_0|} = -{t\ov \th} \ln N + {1\ov 2} - {\th \ov 6} \ .
\ee
For high enough $\th$, a metastability transition occurs at some temperature $t_m$. Indeed, the top line 
in \eqn{ftp22} is positive (it is the entropy) and the $x$-dependent terms in the bottom line will be positive 
provided $3(\th-1) + (1-a) \th x <0$. Since $0<x<a^{-1}$ this implies that for $\th< {3 a\ov 1+ 2 a}\bb=\bb {3\ov N+2}$
the singlet is absolutely stable, otherwise it can become metastable, or eventually unstable (for $\th>1$).
Similarly to the strongly ferromagnetic case, the metastability transition is of first order.

\no
Putting everything together, we obtain the following phase structure:

\no
For $\th >1$ the situation is similar to the strongly ferromagnetic one, leading to tables similar to \ref{table1} and 
\ref{table1a} and the 
symmetry breaking pattern 
\eqn{symmbrea1}.

\no
For ${3\ov N+1} < \th < 1$ the singlet remains stable down to $t=0$, becoming metastable below a
temperature $t_m$, with the one-row state becoming absolutely stable for
$0<t<t_m$. The situation is summarized in table \ref{table3}.

\begin{table}[!ht]
\begin{center}
\begin{tabular}{c@{\hskip 1cm}c}

\begin{tabular}{|c|c|c|c|} \hline
  irrep &  $t<t_m $ & $t_m<t<t_c$ & $ t_c<t$
  \\ \hline \hline
{\rm singlet}      &  {\rm metastable}   &  {\rm stable }  & {\rm stable}     
\\ \hline
 {\rm one-row}    &  {\rm stable} &   {\rm metastable}    &  $\times $    
 \\ \hline
\end{tabular}
\hskip -.5 cm
&

\hskip -.5  cm
\begin{tabular}{|c|c|c|} \hline
    $t<t_m $ & $t_m<t<t_c$ & $ t_c<t$ 
  \\ \hline \hline
 {\rm $\bullet$}  & 
 \rule{0pt}{1.2em}
 \scalebox{0.6}{ \begin{ytableau}
\none {} \hskip -.75 cm  & & & \end{ytableau}} 
 &    
\\
\scalebox{0.6}{\begin{ytableau}
\none {} \hskip -.5 cm  & & &  & \end{ytableau}} &
   {\rm $\bullet$}  &  {\rm $\bullet$}  
 \\ \hline
\end{tabular}

\end{tabular}
\end{center}
\vskip -.5 cm
\caption{\small{${3\ov N+1} < \theta < 1$: Phases for $N\geqslant 3$ and their stability (left); presented according to their free energy as explained in table
\ref{table1a} (right).}}
\label{table3}
\end{table}
\vskip -0.4cm
\no
For ${2\ov N}  < \th < {3\ov N+1} $ the singlet remains absolutely stable at all temperatures.
The situation is summarized in table \ref{table4}.

\begin{figure} [th!] 
\begin{center}
\hskip 2cm\includegraphics[height= 7.5 cm, angle=0]{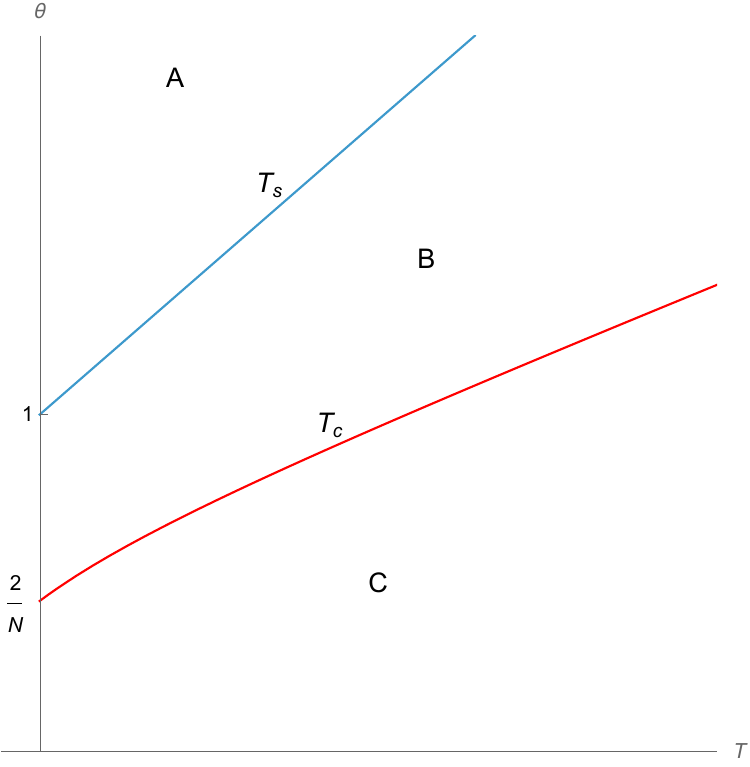}
\end{center}
\vskip -.7 cm
\caption{\small{Qualitative phase diagram of the weakly ferromagnetic model on the $T - \theta$ plane for generic
$N\geqslant 3$.
The temperature is in units
of $|T_0 |$. The straight (green) line $T_s =  \th-1$ is the frontier of stability of the singlet, while the (red) curve
$T_c$ marks a transition to the one-row state.
In region A the state is one-row. In region B a one-row state and the singlet coexist and in region C the state is the singlet. 
Metastability transitions are not depicted.}}
\label{figweak}
\end{figure}

\begin{table}[!ht]
\begin{center}
\begin{tabular}{c@{\hskip 1cm}c}

\begin{tabular}{|c|c|c|} \hline
  irrep & $t<t_c$ & $ t_c<t$
  \\ \hline \hline
  {\rm singlet}  &  {\rm stable}   & {\rm stable}     
  \\ \hline
  {\rm one-row} & {\rm metastable}      &  $\times $     
  \\ \hline
\end{tabular}

\hskip .2 cm 
&

\hskip .2 cm 

\begin{tabular}{|c|c|} \hline
  $t<t_c$ & $t_c<t$ 
  \\ \hline \hline
  \rule{0pt}{1.2em}
  \scalebox{0.6}{
    {\begin{ytableau}
      \none {} \hskip -0.75cm & & & 
    \end{ytableau}}
  } 
  & 
  \\
  {\rm $\bullet$} & {\rm $\bullet$}
  \\ \hline
\end{tabular}

\end{tabular}
\end{center}
\vskip -.5 cm
\caption{\small{${2\ov N} < \theta < {3\ov N+1}$: Phases for $N\geqslant 3$ and their stability (left);
presented according to their free energy as explained in table \ref{table1a} (right).}}
\label{table4}
\end{table}
\vskip-0.4cm
\no
Finally, if $  \th <  {2\ov N}$ the only state is the singlet.
The phase structure of the model is summarized in
figure \ref{figweak}.
\vskip -0.3cm

\newpage
\section{The mixed case $T_0>0$, $\Theta_0< 0 $}
\label{mixed}

This is the most interesting and nontrivial case, and will manifest a very rich phase structure.
Since $\Th_0<0$, we expect that for high enough parameter $\th$ in \eqn{thta} the behavior
will become progressively more paramagnetic. 
Due to the possibility of non-convex behavior for the function $f(x)$, defined in \eqn{jfghj}, in this case
and the fact that, as we shall see,
several different phases may coexist, we call this phase mixed. We will present
its general analysis for arbitrary $N$, and will subsequently fully study the case $N=3$.

\subsection{General analysis}

\begin{figure} [th!] 
\begin{center}
\includegraphics[height= 6.5 cm, angle=0]{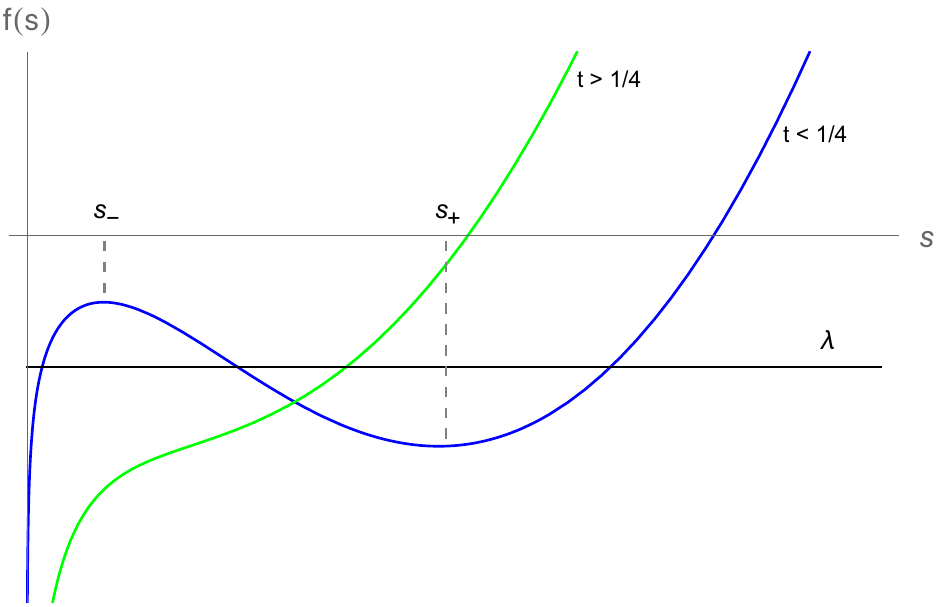}
\end{center}
\vskip -.6 cm
\caption{\small{The function $f(s)$ for $t>{1\ov 4}$ (green) and $t<{1\ov 4}$ (blue). In the first case $f(s)=\lambda$
always has a single solution, while in the the latter it will have three solutions for a range of $\lambda$.}}
\label{2kaboures}
\end{figure}

To efficiently deal with this case, in addition to the rescalings \eqn{thta} it is convenient to also rescale 
the $x_i$ as 
\be
s_i =N\th x_i\ , \qq i= 1,2,\dots, N \ .
\ee
Then \eqn{jfghj} become
\be
\label{jfghj1} 
t \ln s_i - s_i + \ha s_i^2 = \l \ ,\qq i= 1,2,\dots, N\ ,
\ee
for a new Lagrange multiplier $\l$, having shed all parameters other than $t$. This leaner form is useful
when \eqn{jfghj1} has three distinct solutions for $s_i$, as will be the case here, in which case the parametrization
\eqn{xixi1} in terms of a single variable $x$ is not possible. Indeed, defining 
\be
\label{spm}
s_\pm = \ha \pm \sqrt{{1\ov 4} - t }\ ,\qq 0<s_-<\ha \,\qq  \ha < s_+ < 1\ ,
\ee
we see that there can be either one or three solutions $s_1<s_2 < s_3$ to \eqn{jfghj1} 
(see fig.~\ref{2kaboures}):
\be
\label{t14}
\begin{split}
& t>{1\ov 4}: \quad \text{one solution}\ ,\\
& t<{1\ov 4}: \quad \text{one or three solutions depending on $\l$}\ .
\end{split}
\ee
\vskip -.3 cm
\no
So, for $t>{1\ov 4}$ the only state is the singlet. Assuming $t<{1\ov 4}$, the three solutions are ordered as 
\be 
0<s_1<s_- <s_2<s_+<s_3\ .
\label{s1s2s3}\ee
The constraint \eqn{lengyt1} becomes
\be
\label{lengyt2}
\sum_{i=1}^N s_i = N \th\  ,
\ee
this being the only appearance of the parameter $\th$, and the stability coefficients become
\be
C_i = { s_i \ov (s_i-s_+)(s_i-s_-)}\ ,
\ee
clearly satisfying
\be\label{C123}
C_1>0\ ,\qq C_2<0\ ,\qq C_3 >0\ .
\ee
Assuming that $p_1, p_2$ and $p_3$ of the $s_i$ are at solutions $s_1, s_2$ and $s_3$ respectively
(with $p_1+p_2+p_3=N$), stability
requires either $p_2=0$, or $p_2=1$ and $\displaystyle \sum_{i=1}^N C_i <0$.
For the singlet, in which all $s_i =\th$, the stability condition implies
\be
\label{stabat}
\begin{split}
\text{Singlet:}\quad& \th>1:\quad \text{stable for all $t$}.
\\
& \th <  1:\quad {\rm stable\ for}\  t>t_s = \th(1-\th)\,,\ \  {\rm unstable\ for}\ t< t_s \ .
\end{split}
\ee
Eliminating the Lagrange multiplier $\l$ from \eqn{jfghj1} we obtain the system of equations
\be
\label{jfghj2} 
\begin{split}
& t \ln {s_1\ov s_2} - (s_1-s_2) \Big(1 - \ha (s_1+s_2)\Big) = 0 \ ,
\\
& t \ln {s_3\ov s_2} - (s_3-s_2) \Big(1- \ha (s_3+s_2)\Big) = 0 \ ,
\end{split}
\ee
(the corresponding equation for $s_1,s_3$ is redundant), and the constraint \eqn{lengyt2} implies the 
additional equation
\be
\label{pspsps}
p_1 s_1 + p_2 s_2 + p_3 s_3 = \th N\ .
\ee
Solutions of (\ref{jfghj2} and \ref{pspsps}) with $p_2=0$, and thus two different values for $s_i = \{s_1,s_3\}$, 
correspond to states with $p_3$ equal rows in their Young tableau, while solutions with $p_2 =1$ and
thus three different values for $s_i = \{s_1,s_2,s_3\}$ 
correspond to states with $p_3$ equal rows and one smaller row in their Young tableau.

\no
Consider a state with either $p_3$ equal rows $(p_1\bb=\bb N\bb-\bb p_3,p_2\bb=\bb0,p_3)$ or with an extra smaller row
$(p_1\bb=\bb N\bb-\bb1\bb-\bb p_3,p_2\bb=\bb1,p_3)$. At a critical  temperature $t_-$ at which 
$s_1\bb=\bb s_2\bb=\bb s_-$ the states either remain as they are or one transitions to the other. 
Similarly, at the critical temperature $t_+$ at which $s_2\bb=\bb s_3= s_+$, states
$(p_1\bb=\bb N\bb-\bb p_3,p_2\bb=\bb0,p_3)$ and
$(p_1\bb=\bb N\bb-\bb p_3,p_2\bb=\bb1,p_3\bb-\bb1)$ will either remain the same or one will transition to the other.
In both cases, one has to check if the solutions arising at $t=t_\pm$ are stable, and the realization of these
transitions will depend on the value of $\th$.  

\no
The critical temperatures $t_\pm$ can be obtained by inserting $s_1=s_2=s_-$, respectively
$s_2=s_3=s_+$, in \eqn{jfghj2} and \eqn{pspsps}. We obtain the transcendental equations
\be
\label{trans1}
 t_\pm \ln {s_\pm\ov s_{3\pm}} - (s_\pm -s_{3\pm}) \Big(1 - \ha (s_\pm + s_{3\pm})\Big) = 0 \ ,\qquad s_{3\pm} =  s_\pm + (\th-s_\pm){N\ov p_3}\ ,
\ee
with $s_\pm$ given by \eqn{spm} 
(in the $s_2=s_3=s_+$ case we reversed the notation from $s_1,p_1$ to $s_3,p_3$ and vice versa, thus
naming $s_1,s_2$ the two coalescing solutions and $s_3$ the remaining one in either $t_-$ or $t_+$).

\no
Equations \eqn{trans1} admit the analytical solution
\be
\label{anal} 
\begin{split}
\th < \half :&\qquad t_- = t_s = \th(1-\th) \ ,
\\
\half <\th < 1 :&\qquad t_+ = t_s = \th(1-\th) \ .
\end{split}
\ee
In this temperature, $s_1= s_2 = s_3 = \th$, and thus the state for any $(p_1,p_2,p_3)$
degenerates to the singlet. This is also the temperature above which the singlet becomes stable according to \eqn{stabat}.
Therefore, $t_s$ marks transitions from general states to the singlet.
Note that $p_2$ does not enter in the equations \eqn{trans1}. However, its particular value 
$p_2 = 0,1$ affects the stability of the solutions as $t$ approaches $t_\pm$.

\no
The existence of values of $t_+$ and $t_-$ other than $t_s$
depend on the value of $\th$ relative to a set of critical values; specifically (not necessarily in order)
\be
\label{514}
\th_0 =1-{p_3\ov N} \ ,\quad \th_1 = {2p_3 \ov N} ,\quad \th_2 = {2(N-p_3) \ov N}\ ,\quad
\th_c \ , \quad {1\over 2}\ , \quad 1\ .
\ee
The critical value $\th_c$ is determined by setting both \eqn{trans1}
and its derivative with respect to $t_-$ to zero, or the corresponding conditions for $t_+$ (these conditions
for $t_-$ and $t_+$ are never simultaneously satisfied).
The relative ordering of the various critical $\th$ values determines the qualitative properties of the phase
flow. 

\no
For $\th>1$, equations \eqn{trans1} admit no acceptable solutions.
 For $\th<1$, denoting $t_1,t_2$
generic transcendental solutions of \eqn{trans1}, the corresponding critical temperatures
in each range of $N/p_3$ and $\theta$ are given below:
\ba
{N\over p_3}< 2: \qquad\qquad \th<\th_c :& &t_- = t_s \ , \quad \text{no~~} t_+\cr
 \quad \th_c < \th <\th_0:& &t_-=t_s\ , \quad t_+ = t_1 , ~t_2\cr
 \quad \th_0 < \th <\textstyle{1\over 2}:& &t_-=t_s\ , \quad t_+ = t_1 \cr
 \quad \textstyle{1\over 2} < \th < 1:& &t_-=t_1\ , \quad t_+ = t_s \cr & &\cr
2<{N\over p_3}< 3: \qquad\qquad \th<\th_c:& &t_- = t_s \ , \quad \text{no~~} t_+\cr
 \quad \th_c < \th <\textstyle{1\over 2}:& &t_-=t_s\ , \quad t_+ = t_1 , ~t_2\cr
 \quad \textstyle{1\over 2} < \th <\th_0 :& &t_-=t_1\ , \quad t_+ =t_s , ~ t_2 \cr
 \quad \th_0 < \th < \th_1:& &t_-=t_1\ , \quad t_+ = t_s \cr
 \quad \th_1 < \th < 1:& & t_+ = t_s\ , \quad \text{no}~~t_- \cr & &
 \label{rangerover}
 \\
 3<{N\over p_3}< 4: \qquad\qquad  \th<\th_c:& &t_- = t_s \ , \quad \text{no~~} t_+\cr
 \quad \th_c < \th <\textstyle{1\over 2}:& &t_-=t_s, t_1 , t_2\ , \quad \text{no}~~t_+\cr
 \quad \textstyle{1\over 2} < \th <\th_1 :& &t_-=t_1\ , \quad t_+ =t_s , ~ t_2 \cr
 \quad \th_1 < \th < \th_0:& &t_+=t_s , t_1\ , \quad \text{no}~~ t_- \cr
 \quad \th_0 < \th < 1:& & t_+ = t_s\ , \quad \text{no}~~t_- \cr & &\cr
 4<{N\over p_3}: \qquad\qquad  \th<\th_c:& &t_- = t_s \ , \quad \text{no~~} t_+\cr
 \quad \th_c < \th <\th_1:& &t_-=t_s, t_1 , t_2\ , \quad \text{no}~~t_+\cr
 \quad \th_1< \th <\textstyle{1\over 2}:& &t_-=t_s, t_1\ ,\quad \text{no}~~t_+ \cr
 \quad \textstyle{1\ov 2} < \th < \th_0:& &t_+=t_s , t_1\ , \quad \text{no}~~ t_- \cr
 \quad \th_0 < \th < 1:& & t_+ = t_s\ , \quad \text{no}~~t_- \nonumber
 \ea
 At the transition value $N/p_3 = 2$, $\th_0 = {1\ov 2},\ \th_1 =1$; for $N/p_3 = 3$,
 $\th_c = {1\ov 2}, \ \th_0 = \th_1$; and for $N/p_3 = 4$, $\th_1 = {1\ov 2}$.
 The temperatures $t_\pm$ for these values are obtained
 by eliminating the intervals of $\th$ that shrink to zero.

\no
Critical temperatures $t_\pm$ other than $t_s$, marking general transitions, are determined
by nontrivial solutions of \eqn{trans1} and are accessible only numerically. If they correspond to a transition
involving a state with a number of equal rows ($p_2=0$), $p_1 , p_3$ can be arbitrary.
If, however, they correspond to a transition of a state with one unequal row ($p_2=1$), then 
the stability of solutions to \eqn{jfghj2} infinitesimally close to $s_1=s_2=s_-$  and to $s_2=s_1=s_+$ (recall 
the renaming described below \eqn{trans1}) requires $p_1=1$ (and thus $p_3=N-2$) so that the divergences
$C_2 \simeq -C_1 \to +\infty$ cancel out in the stability condition $C_1+C_2+(N-2) C_3 <0$ near $t = t_\pm$.
The finite residue in $C_1 + C_2$ turns out to be negative, and should prevail over the positive term $(N-2) C_3$.
Numerically, we found that stability also requires  $N-2=1$, leaving $N=3$ as the only possibility.
For this reason we will focus on the case of $SU(3)$ in the next subsection.

 \no
 Other critical temperatures, in addition to $t_\pm$, may arise when solutions to (\ref{jfghj2} and \ref{pspsps})
 cease to exist. Further, free energy comparison of coexisting stable states will introduce transition temperatures
 in which their metastability properties are swapped.



\no
For $p_2=0$, the $s_i$ (or $x_i$) can have only two distinct values, and we can use the parametrization
\eqn{xixi1} in terms of a single $x$ with $p = p_3$. The corresponding equation \eqn{gh} becomes
\be
\label{gh2}
t \ln {1+x\ov 1-a x} -\th (1+a) x \Big(1 -\th \big(1+{1-a\ov 2} x\big)\Big) = 0 \ , \quad \text{where}~~a= {p_3\ov N-p_3} \ .
\ee
Note that in the present case there is no restriction on $p_3$ (unlike in previous cases), since both
$s_1 = N\th x_1$ and $s_3 = N\th x_3$ correspond to positive $C_i$, as per
\eqn{C123}. Solutions of \eqn{gh2} with $x>0$ correspond to states with $p_3$ equal rows, and solutions
with $x<0$ to states with $N-p_3$ equal rows.

\no
A state will cease to exist at a 
critical temperature parameter $t_c$, determined by setting, both the left hand side of \eqn{gh2} and its $t$-derivative
to zero, as in previous sections. Combining the two equations gives
\be
\label{tcgh2}
t_c = \th (1+x_c)(1-a x_c)\Big(1-\th \big(1+(1-a)x_c\big)\Big)\ ,
\ee
which expresses $t_c$ in terms of the critical parameter $x_c$ at the transition. 
Substituting the above expression back to \eqn{gh2} gives a transcendental equation for $x_c$ that can be solved numerically,
and in turn determines $t_c$.

\no
At particular values of $\th$ the critical temperature $t_c$ vanishes, determining the range
in which such transitions can occur. From \eqn{tcgh2}, $t_c$ vanishes for $x=-1$, $x=1/a$ and 
$x={1-\th\ov \th(1-a)}$, and from \eqn{gh2} we find the respective critical values (in agreement with \eqn{514})
\be
\label{fhuh9}
\th_1= {2a\ov 1+ a} = {2p_3\ov N}\ ,\quad \th_2= {2 \ov 1+ a} = {2(N-p_3)\ov N}\ , \quad \th_0=1 \ .
\ee
The stability coefficients \eqn{fluuc} take the form 
\be
\label{cicib}
\begin{split}
& C_i^{-1} = N T_0 \Big( {t \ov \th( 1+ x)} - 1 + \th (1+x)\Big)   \, ,\qquad i=1,2,\dots, p_3\ ,
\\
& C_i^{-1} =  N T_0 \Big( {t \ov \th( 1- a x)} - 1 + \th (1-a x)\Big)  \,  ,\quad i=p_3+1,2,\dots, N\ .
\end{split}
\ee

\no
At low temperature the solution of \eqn{gh2} can approach either $x=1/a$ or $x=-1$, corresponding to states
with either $p_3$ or $N-p_3$ equal rows of maximal length:
\be
\begin{split}
x = &{1\ov a} \bigg(1- {2\ov \th_1} e^{-b_1/t}+\dots  \bigg) , 
\quad b_1 = 2{\th\ov\th_1} \Big(1-{\th\ov\th_1}\Big)\ , \quad \text{for}~~ \th < \th_1 \ , \\
x = &-1+ {2\ov\th_2} e^{-b_2/t}+\dots \, ,\qquad  b_2 = 2{\th\ov\th_2} \Big(1-{\th\ov\th_2}\Big)\ ,
\quad \text{for}~~ \th < \th_2 \ .
\end{split}
\ee
The two expressions are related by $p_3 \to N-p_3$, consistent with \eqn{symaa}.

\no
The above limiting solutions are not necessarily stable.
Of particular interest are the stable configurations at low temperature. A detailed analysis reveals that, at temperature
near zero, the locally stable configurations are the singlet, if $\th>1$, the one-row configuration, if $\th < {2\ov N}$,
and the $p$-equal rows configuration ($2 \leqslant p\leqslant N-1$), if ${p\ov N} < \th < {2 p \ov N}$.
This means that 
several phases can coexist. For instance, for $N=3$, the singlet and two-row
will coexist for $1\leqslant \th \leqslant {4\ov 3}$;
 for $N=4$, the two-row and the three-row with coexist for ${3\ov 4} \leqslant  \th \leqslant 1$ and
 the singlet and three-row will coexist for $1\leqslant \th\leqslant {3\ov 2}$;  
and for $N=5$, the two-row and three-row will coexist for
${3\ov 5}\leqslant\th\leqslant{4\ov 5}$, the three-row and four-row will coexist for ${4\ov 5}\leqslant\th\leqslant 1$,
and the singlet, three-row, and four-row will coexist for $1\leqslant\th\leqslant{6\ov 5}$, in the first instance of
three phases coexisting. 
In this latter case, the $SU(5)$ symmetry group is either unbroken in the singlet state, or
broken to $SU(3)\times SU(2)\times U(1)$ in the three-row state or $SU(4)\times U(1)$ in the four-row state.
Also, for ${6\ov 5} \leqslant\th\leqslant{8 \ov 5}$ the singlet and the four-row states coexist.  
The metastability properties of coexisting phases are resolved in each case by examining their free energy.

\subsection{The $SU(3)$ case}

The analysis of the previous section makes clear that the phase diagram of this model is quite involved.
Moreover, the cases of $SU(3)$ and $SU(4)$ are qualitatively different from $SU(N)$ with $N>4$,
as is clear from the ranges of $N/p_3$ in \eqn{rangerover}. The case of $SU(3)$, in particular, presents the additional
unique feature of having states with two unequal rows, as discussed earlier, and deserves special study.

\no
We will analyze below the case of $SU(3)$. Nontrivial states
correspond to symmetric single-row irreps, doubly symmetric irreps with two equal rows, and states
with two unequal rows. Symmetric states correspond to $p_2=0$, $p_3=1$ and doubly symmetric ones
to $p_2=0$, $p_3=2$, and both are described by the single equation \eqn{gh2}. The general state with two
unequal rows corresponds to $p_1=p_2 =p_3=1$ and its critical temperatures are as in \eqn{rangerover}
for $N/p_3=3$, with $\th_0 = \th_1 = {2\ov3}$ and $\th_2 = {4\ov 3}$.

\no
With the exception of the transition temperature $t_s$, the remaining analysis is done numerically.
It will involve a nontrivial critical temperature $t_c$, given by \eqn{tcgh2}
for $p_2=0$ and by $t_\pm$ in (\ref{trans1},\ref{rangerover}) for $p_2=1$. 
Resolution of metastability when more than one stable states coexist requires comparison of their
free energies, and will bring in additional critical temperatures and values of $\th$.
We omit the details and present the results for the phase structure and transitions of the system
within various ranges of $\th$. States with two unequal rows are denoted as 1+1 rows. For completeness
we also include unstable states, although they are physically irrelevant. The symbol $\times$
means that the state is absent.

\no
\underline{$\th<{1\ov 2}$:}
\vskip -1cm
\begin{table}[!ht]
\begin{center}
\begin{tabular}{|c|c|c|c|c|} \hline
  irrep & $t<t_s$ & $t_s<t<t_m$ & $t_m<t<t_c$ & $t_c<t$ 
  \\ \hline 
  {\rm singlet}  & {\rm unstable}     & {\rm metastable} & {\rm stable}   & {\rm stable} \\ \hline
  {\rm one-row}  & {\rm stable}       & {\rm stable}     & {\rm metastable} & $\times$   \\ \hline
\end{tabular}
\end{center}
\vskip -0.5cm
\caption{\small{Phases for $N= 3$ in the regime $ \theta < \ha $ and their stability.}}
\label{table5}
\end{table}
\vskip -0.3cm
\begin{table}[!ht]
\begin{center}
\begin{tabular}{|c|c|c|c|} \hline
  $t<t_s$ & $t_s<t<t_m$ & $t_m<t<t_c$ & $t_c<t$ \\ \hline \hline
 & 
  \rule{0pt}{1.5em}
  {\rm $\bullet$}   &  \scalebox{0.6}{
    \begin{ytableau}
      \none {} \hskip -0.65cm & & & 
    \end{ytableau}
  }  &
  \\
  \rule{0pt}{1.2em}
  \scalebox{0.6}{
    \begin{ytableau}
      \none {} \hskip -0.65cm & & & & &
    \end{ytableau}
  } 
  & \scalebox{0.6}{
    \begin{ytableau}
      \none {} \hskip -0.7cm & & & & 
    \end{ytableau}
  } 
& {\rm $\bullet$} & {\rm $\bullet$}
  \\ \hline
\end{tabular}
\end{center}
\vskip -0.5cm
\caption{\small{Phases for $N= 3$ in the same regime $ \theta < \ha $ presented according to their free energy.}}
\label{table6}
\end{table}

\newpage
\no
\underline{$\ha < \th<{2\ov 3}$:}

\begin{table}[!ht]
\begin{center}
\begin{tabular}{|c|c|c|c|c|c|c|} \hline
  irrep & $t<t_-$ & $t_-<t<t_+$ & $t_+<t<t_s$ & $t_s<t<t_m$ & $t_m<t<t_c$ & $t_c<t$ 
  \\ \hline 
  {\rm singlet}    & {\rm unstable}   & {\rm unstable}   & {\rm unstable}   & {\rm metastable} & {\rm stable}    & {\rm stable} \\ \hline
  {\rm one row}    & {\rm stable}     & {\rm unstable}   & {\rm unstable}   & $\times$         & $\times$        & $\times$     \\ \hline
  {\rm 1+1 rows}   & $\times$         & {\rm stable}     & $\times$         & $\times$         & $\times$        & $\times$     \\ \hline
  {\rm two rows}   & {\rm unstable}   & {\rm unstable}   & {\rm stable}     & {\rm stable}     & {\rm metastable} & $\times$     \\ \hline
\end{tabular}
\end{center}
\vskip -0.4cm
\caption{\small{Phases for $N= 3$ in the regime ${1\over 2} < \theta < {2\over 3}$ and their stability.}}
\label{table7}
\end{table}
\vskip -0.3cm
\begin{table}[!ht]
\begin{center}
\begin{tabular}{|c|c|c|c|c|c|} \hline
$t<t_-$ & $t_-<t<t_+$ & $t_+<t<t_s$ & $t_s<t<t_m$ & $t_m<t<t_c$ & $t_c<t$ 
\\ \hline \hline

& & & {\rm $\bullet$}& \rule{0pt}{1.2em}  \scalebox{0.6}{
    \begin{ytableau}
      \none {} \hskip -0.85cm & & & 
      \\
      \none {} \hskip -0.85cm & &  &
    \end{ytableau} 
  }  & \\

 
 \scalebox{0.6}{
    \begin{ytableau}
      \none {} \hskip -0.5cm & & & & & & 
    \end{ytableau}
  }   & \scalebox{0.6}{
    \begin{ytableau}
      \none {} \hskip -0.5cm & & & & & 
      \\ 
      \none {} \hskip -0.5cm & & & 
    \end{ytableau} 
  }   &  \scalebox{0.6}{
    \begin{ytableau}
      \none {} \hskip -0.5cm & & & & & 
      \\
      \none {} \hskip -0.5cm & &  & & &
    \end{ytableau} 
  } &   \scalebox{0.6}{
    \begin{ytableau}
      \none {} \hskip -0.7cm & & & & 
      \\
      \none {} \hskip -0.7cm & &  & &
    \end{ytableau} 
  } &  {\rm $\bullet$} & {\rm $\bullet$}    \\[2ex] \hline

\end{tabular}
\end{center}
\vskip -0.4cm
\caption{\small{Phases for $N= 3$ in the same regime  ${1\over 2} < \theta < {2\over 3}$ according to their free energy and following the structure in 
table \ref{table7}.}}
\label{table8}
\end{table}



\no
\underline{${2\ov 3}<\th<1 $:}
\vskip -0.6cm

\begin{table}[!ht]
\begin{center}
\begin{tabular}{|c|c|c|c|c|} \hline
  irrep & $t<t_s$ & $t_s<t<t_m$ & $t_m<t<t_c$ & $t_c<t$ 
  \\ \hline 
  {\rm singlet}  & {\rm unstable}     & {\rm metastable} & {\rm stable}   & {\rm stable} \\ \hline
  {\rm two-rows}  & {\rm stable}       & {\rm stable}     & {\rm metastable} & $\times$   \\ \hline
\end{tabular}
\end{center}
\vskip -0.3cm
\caption{\small{Phases for $N= 3$ in the regime ${2\over 3} < \theta < 1$ and their stability.}}
\label{table9}
\end{table}
\vskip -0.3cm

\begin{table}[!ht]
\begin{center}
\begin{tabular}{|c|c|c|c|} \hline
  $t<t_s$ & $t_s<t<t_m$ & $t_m<t<t_c$ & $t_c<t$ \\ \hline \hline
 & 
  \rule{0pt}{1.5em}
  {\rm $\bullet$}   &  \scalebox{0.6}{
    \begin{ytableau}
   \hskip -0.5cm & & 
      \\  & &  
    \end{ytableau}
  }  &
  \\
  \rule{0pt}{1.2em}
  \scalebox{0.6}{
    \begin{ytableau}
    \hskip -0.5cm & & &  &
    \\ & & &  &
    \end{ytableau}
  } 
  & \scalebox{0.6}{
    \begin{ytableau}
\hskip -0.5cm & & &  
      \\ & & &  
    \end{ytableau}
  } 
& {\rm $\bullet$} & {\rm $\bullet$}
  \\[2ex] \hline
\end{tabular}
\end{center}
\vskip -0.4cm
\caption{\small{Phases for $N= 3$ in the same regime  ${2\over 3} < \theta < 1$ according to their free energy and following the structure in 
table \ref{table9}.}}
\label{table10}
\end{table}

\underline{$1<\th<{6\ov 5} $:} 
\begin{table}[!ht]
\begin{center}
\begin{tabular}{c@{\hskip 1cm}c}

\begin{tabular}{|c|c|c|c|} \hline
  irrep &  $t<t_m $ & $t_m<t<t_c$ & $ t_c<t$
  \\ \hline \hline
{\rm singlet}      &  {\rm metastable}   &  {\rm stable }  & {\rm stable}     
\\ \hline
 {\rm two-rows}    &  {\rm stable} &   {\rm metastable}    &  $\times $    
 \\ \hline
\end{tabular}
\hskip -.5 cm
&

\hskip -.5  cm
\begin{tabular}{|c|c|c|} \hline
    $t<t_m $ & $t_m<t<t_c$ & $ t_c<t$ 
  \\ \hline \hline
 {\rm $\bullet$}  & 
 \rule{0pt}{1.4em}
 \scalebox{0.57}{ \begin{ytableau}
 \none{} \hskip -0.75cm & & & 
\\
\none{} \hskip -0.75cm & & &   \end{ytableau}} 
 &    
\\
\scalebox{0.6}{\begin{ytableau}
 \hskip -.5 cm  & & &  & 
\\
 & & &  & \end{ytableau}} & \rule{0pt}{1.1em}
   {\rm $\bullet$}  &  {\rm $\bullet$}  
 \\[2ex]\hline
\end{tabular}

\end{tabular}
\end{center}
\vskip -.4 cm
\caption{\small{Phases for $N= 3$ in the regime $1<\th <{6\ov 5}$ and their stability (left); presented according to their free energy (right).}}
\label{table11}
\end{table}

\newpage
\no
\underline{${6\ov 5}<\th< {4\ov 3}$:}

\begin{table}[!ht]
\begin{center}
\begin{tabular}{c@{\hskip 1cm}c}

\begin{tabular}{|c|c|c|} \hline
  irrep & $t<t_c$ & $ t_c<t$
  \\ \hline \hline
  {\rm singlet}  &  {\rm stable}   & {\rm stable}     
  \\ \hline
  {\rm two-rows} & {\rm metastable}      &  $\times $     
  \\ \hline
\end{tabular}

\hskip .2 cm 
&

\hskip .2 cm 

\begin{tabular}{|c|c|} \hline
  $t<t_c$ & $t_c<t$ 
  \\ \hline \hline
  \rule{0pt}{1.2em}
  \scalebox{0.6}{
    \begin{ytableau}
  \none{}  \hskip -0.8cm & & &
\\
  \none{}  \hskip -0.8cm         & & &
    \end{ytableau}
  } 
  & 
  \\
  {\rm $\bullet$} & {\rm $\bullet$}
  \\ \hline
\end{tabular}

\end{tabular}
\end{center}
\vskip -.4 cm
\caption{\small{Phases for $N= 3$ in the regime ${6\ov 5} <\th <{4\ov 3}$  and their stability (left); presented according to their free energy (right).}}
\label{table12}
\end{table}

\no
\underline{$\th >{4\ov 3}$:}\quad  the only state is the singlet.

\begin{figure} [th!] 
\begin{center}
\includegraphics[height= 9.5 cm, angle=0]{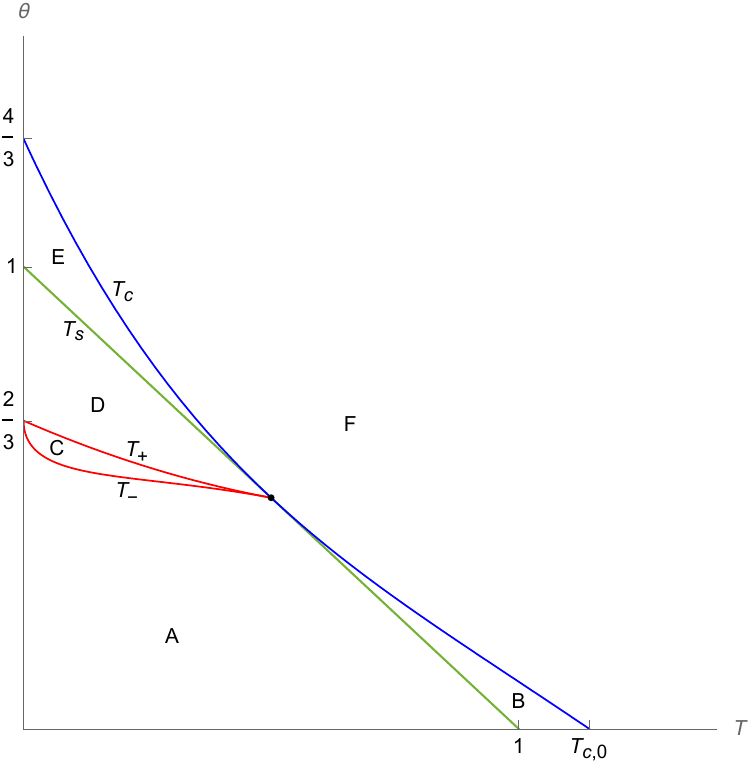}
\end{center}
\vskip -.6 cm
\caption{\small{Qualitative phase diagram of the mixed model on the $T - \theta$ plane. The temperature is in units
of $T_0$. The straight (green) line $T_s = 1-\theta$ is the frontier of stability of the singlet, while the (blue and red) curves
$T_c$ and $T_\pm$ mark transitions of nontrivial representations.
In region A the state is one-row. In region B a one-row state and the singlet coexist. In region C the state 
has two unequal rows. In region D the state has two equal rows. In region E a state with two equal rows and the singlet
coexist. Finally, in region F the state is the singlet. The point $T = \theta = \half$ where all curves meet is a multiple critical point.
Metastability transitions are not depicted.}}
\label{figmix}
\end{figure}

\no
The phase structure of the model
in the $T - \theta$ plane is summarized in figure \ref{figmix}.

\no
We see that the system has markedly different properties depending on the value of $\th$.
For $\th<1/2$ it becomes purely ferromagnetic, with a phase flow qualitatively similar to that of the
strongly ferromagnetic case. The intermediate range $1/2<\th<2/3$ is the most interesting one, and the
only one in which phases with two unequal rows appear, along with simply and doubly symmetric ones.
The behavior for $2/3<\th<1$ is similar to the
standard ferromagnetic one, with the crucial difference that, now, the magnetized state is a doubly
symmetric one, and for $\th>2/3$ the symmetric one never reappears. For $\th>1$ the singlet becomes
stable all the way down to zero temperature, and absolutely stable for $\th>6/5$. Finally,  for $\th>4/3$
the system becomes paramagnetic with the singlet as the only state.

\section{Conclusions}
\label{conclusions}

The addition of three-body interactions in the $SU(N)$ fundamental ferromagnet adds new elements and modifies
their phase structure. In the mean field approximation the system has two relevant parameters,
the strengths of two- and three- body interactions, and one dimensionless parameter, their relative strength, that
alters and determines their qualitative behavior. Interestingly, {\it anti}ferromagnetic three-body interactions coupled with
ferromagnetic two-body ones give rise to a novel phase which breaks the global $SU(N)$
symmetry and has two distinct order parameters. All this in addition
to overlapping phases, hysteresis, and metastability transitions that are the hallmark of $SU(N)$
ferromagnetism. In particular, for the $SU(3)$ ferromagnet the  symmetry breaking pattern from high to low temperature is
$SU(3) \to SU(2)\times U(1) \to U(1)\times U(1) \to U(1) \times SU(2)$.

\no
The results in this paper can be generalized along several possible directions. Adding four-body interactions is one
of them, but perhaps not the most relevant one, as we expect higher-body interactions to become progressively
weaker. On the other hand, adding three-body interactions to ferromagnets with atoms in higher representations,
using the techniques developed in \cite{PShigher}, is a promising direction of investigation.
Atoms in irreps of dimension $M$ are still invariant under a global $SU(N)$, with $N<M$, and offer
a systematic way to study systems of reduced symmetry. It will be interesting to explore the phase structure of such
ferromagnets in the presence of three-body interactions. The case of atoms in the fully antisymmetric irrep of
$SU(N)$ of dimension $N(N-1)/2$ is particularly intriguing, since the system already manifests a rich phase structure
with a singlet, one-row, and two-row phases overlapping for a range of temperatures
as metastable states. It will be interesting to see how the addition of three-body interaction enriches or modifies
the phase flow.

\no
The response of the ferromagnet to the application of external $SU(N)$ magnetic fields is another interesting issue.
The full phase diagram of the model with a magnetic field in a single Cartan direction
is already quite intricate \cite{PSferro}, and the addition of three-body interactions is expected to introduce additional
phases and complexity.
Further, the relevance of our results to topological phases of nonabelian models, as have been
proposed in one dimension
\cite{RQ,CFLT,TLC,RPAR}, and the persistence of such phases in higher dimensions are interesting topics for exploration.

\no
Finally, the realization of ferromagnets in physical systems such as cold atoms and the experimental verification
of the results of the present paper and of previous related work remain the most interesting and physically relevant
open issues.

\subsection*{Acknowledgements }

The research of A.P. was supported by the National Science Foundation 
under grants NSF-PHY-2112729 and NFS-PHY-2112479, and by PSC-CUNY grants 67100-00 55 and 6D136-00 04.\\
K.S. would like to thank the Department of Theoretical Physics at CERN for financial support and hospitality during the
late stages of this research.


\end{document}